\documentclass[english,prl,twocolumn,showkeys,showpacs,floatfix]{revtex4}
\usepackage[T1]{fontenc} \usepackage[latin1]{inputenc}
\usepackage{babel} \usepackage{epsfig}

\newcommand{\tde}[1]{\dot}

\renewcommand{\Re}{\text{Re\,}}

\begin{document}

\title{Stochastic resonance in a non Markovian discrete state model
  for excitable systems} 
\author{T.~Prager and L.~Schimansky-Geier} \affiliation{Institute of
Physics, Humboldt-University of Berlin, Newtonstr. 15, 12489
Berlin, Germany } \date{\today}

\begin{abstract}
  We study a non Markovian three state model, subjected to an
  external periodic signal.  This model is intended to describe an
  excitable systems with periodical driving.  In the limit of a small
  amplitude of the external signal we derive expressions for the
  spectral power amplification and the signal to noise ratio as well
  as for the inter-spike interval distribution.
\end{abstract}

 \maketitle

%\section{Introduction}
Stochastic resonance (SR) is one of the interesting effects where
noise plays a constructive role in nature. It describes the optimal
response of a nonlinear system to a periodic signal at a finite non
zero noise level.  While today most research in the field of SR is
focused on applications in information processing systems like
neurons, the effect was originally discovered in the context of
climatology \cite{benzi}, where it allowed to explain the periodic
occurrence of ice ages.  

A successful theory of SR in bistable systems was introduced in
\cite{wiesenfeld}.  There a reduction to a two state Markovian dynamic
has been proven to be very useful to obtain analytical expressions of
the signal to noise ratio (SNR),
% which is a
%measure to quantify the quality of the response of a stochastic system
%to an external signal. 
which as a hallmark of SR shows a maximum at a certain finite noise
level.  Another example of a two state theory for a bistable system
with delay has been developed in \cite{pikovsky}. In this system the
spectral power amplification (SPA) shows several maxima as a function
of the frequency.  Array enhanced SR in coupled bistable systems is
studied in \cite{siewert}.  Recently, in \cite{goychuk} a theory of SR
in non Markovian two state models has been established and applied to
ion channel gating dynamics. In
\cite{lindner} a piecewise linear FitzHugh-Nagumo system was modeled
as a two state dynamics which allowed to calculate the SPA 
in this particular case.

%Instead of starting  from a specific
%continuous markovian dynamics and reducing it to a non markovian discrete
%state dynamics, in this paper we follow a different approach,
%mapping the characteristic parts of

In this paper study SR in excitable systems
\cite{longtin,wiesenfeld2,jung,alarcon,marino}
by mapping the characteristic parts of an excitable
stochastic dynamics onto a non Markovian three state dynamics
\cite{prager}.  The resulting model allows to derive 
formulas for relevant quantities like SNR or SPA, which may help to
explain the SR found in real world excitable systems driven by
periodic signals, ranging from crayfish mechanoreceptors
\cite{wiesenfeld2} to semiconductor lasers with optical feedback
\cite{marino}.

A key feature of excitable stochastic systems is a stable fixed point
or rest state from which the system can be excited by noise onto an
excitation loop.  This fixed point corresponds to state $1$ in our
model. The transition out of this state is assumed to be a rate
process, describing the excitation over a threshold due to noise.  The
other two states model two parts of the excitation loop each having a
different output.  One might for example think of these two parts as
the firing and refractory state in the dynamics of an excitable
FitzHugh-Nagumo neuron where the output is high in the firing and low
in the refractory state. The transitions from state $2$ to $3$ and $3$
back to $1$ are governed by arbitrary waiting time distributions
(WTD), which must be adapted to the system to be described by this
model.

The external periodical driving acts onto the system as a periodical
modulation of the rate for the transition $1\to 2$.  The times spent
in the firing and refractory state are assumed not to be affected by
the external driving, which is true for example in a stochastic
FitzHugh-Nagumo system with small driving amplitudes and sufficiently
low driving frequencies.

%Together this three state approach yields a non markovian model of an
%excitable system.

\begin{figure}[htbp]
\centerline{
    \includegraphics[scale=1.0]{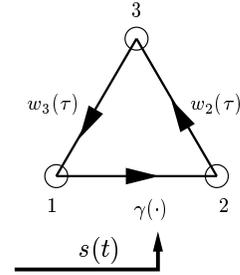}
}
%\hspace{0.5cm}
%  \begin{minipage}[b]{4cm}
\caption{\label{model}
  Three state model of an excitable system. The transition from state
  1 to 2 is governed by a Poisson process whose rate
  depends on the external  signal $s(t)$ . The transitions
  from 2 to 3 and 3 to 1 are governed by arbitrary WTDs
  $w_2(\tau)$ and $w_3(\tau)$, which do not depend on the
  external signal. }
%\end{minipage}
\end{figure}

The state of our model is described by the three conditioned probabilities
$P_{i}(t|t_0),\;i=1,2,3$ that the system is in state $i$ at
time $t$ given
that it has been in state 1 at time $t_0$.
These probabilities obey the generalized master equations
\begin{eqnarray}\label{condprob}
  \tde{t}  P_1(t|t_0)&\!\!=\!\!&\int_{t_0}^t \!\!\!\!d\tau
  \gamma(\tau)P_1(\tau|t_0) (w_2\circ w_3)(t-\tau)-\gamma(t)P_1(t|t_0)
  \nonumber\\ 
  \tde{t}  P_2(t|t_0)&\!\!=\!\!&\gamma(t)P_1(t|t_0)-\int_{t_0}^t \!\!\!\!d\tau
  \gamma(\tau)P_1(\tau|t_0) w_2(t-\tau)\\
  \tde{t}  P_3(t|t_0)&\!\!=\!\!&\int_{t_0}^t \!\!\!\!d\tau
  \gamma(\tau)P_1(\tau|t_0)\big[w_2(t-\tau)-(w_2\circ
  w_3)(t-\tau)\big]
  \nonumber
\end{eqnarray}
with initial condition $P_1(t_0|t_0)=1$, $P_2(t_0|t_0)=0$, $P_3(t_0|t_0)=0$
and $(w_2\circ w_3)(t):=\int_0^t d\tau w_2(\tau) w_3(t-\tau)$.\\
The meaning of eqs.(\ref{condprob}) will be exemplarily explained 
by means of the first one: 
The change in probability to be in  state 1 
is the probability influx minus the probability outflux. 
The later is the instantaneous rate $\gamma(t)$
times the occupation probability $P_1(t|t_0)$,
as the transition from state 1 to state 2 is a Poisson process with
rate $\gamma(t)$.
The influx into state 1 is related to the outflux of state 1
in the past: Consider a
system which has left state 1 at some instant $\tau$ in the
past, i.e.  between $t_0$, when the system has started in state 1 
and the current time $t$.
If this system waits the time
$t-\tau$ in state 2 and 3, which happens with probability 
$(w_2\circ w_3)(t-\tau)$), it will reenter state 1 at time $t$,  
therefore contributing to the influx into state 1 at time $t$.
Taken together this leads to the presented influx term.  
The equations for $P_2(t|t_0)$ and $P_3(t|t_0)$ follow from similar
considerations.

In the case of a constant rate 
those equations can be shown to be equivalent to so called semi Markovian
master equations used to describe continuous time random walks
\cite{kenkre}. 

For the time dependent rate we use a Kramers type rate, which is
modulated by the external signal,
%\begin{eqnarray}\label{rate} 
$\gamma(t)=r_0(D)\exp(-\frac{A_0}{D}\cos{\Omega t})$,
%\end{eqnarray}
where $D$ is the noise strength, $r_0(D)\propto\exp(-\Delta U_{\text{eff}}/
D)$ is the rate without driving
and $A_0$ and $\Omega$ are the effective amplitude
and frequency of the driving.

%\section{Signatures of stochastic resonance}
In order to show the effect of SR in our model we
calculate the SNR and the SPA for small $\alpha:=A_0/D$. 
We assign a high output $x(t)=x_1$ to the system if it is in state 2(firing),
and a low output $x(t)=x_0$ if it is in state 1(rest) and 3(refractory)
respectively.
The spectral power of the output $x(t)$ at the input
frequency $\Omega$ can be derived from the asymptotic mean value
$\left<x(t)\right>_{\text{asy}}=x_0(
P_1(t)_{\text{asy}}+P_3(t)_{\text{asy}})+x_1 P_2(t)_{\text{asy}}$
where the asymptotic probabilities  $P_i(t)_{\text{asy}}
:=\lim_{t_0\to -\infty} P_i(t|t_0)$.
are governed by 
\begin{eqnarray}\label{asympprob}
  \tde{t}  P_1(t)_{\text{asy}}&=&-\gamma(t)P_1(t)_{\text{asy}}+
  \\&&
  \!\!\!\!\int_0^\infty \!\!\!\! d \tau \gamma(t-\tau) P_1(t-\tau)_{\text{asy}}
  (w_2\circ w_3)(\tau)\nonumber
\end{eqnarray}
together with the normalization condition
$P_1(t)_{\text{asy}}+ P_2(t)_{\text{asy}}+ P_3(t)_{\text{asy}}=1$ where
\begin{eqnarray}
  P_2(t)_{\text{asy}}&=&\int_0^\infty  d\tau \gamma(t-\tau
  )P_1(t-\tau)_ {\text{asy}}
  z_2(\tau)\\
  P_3(t)_{\text{asy}}&=&\\
&&\hspace{-1.5cm}\int_0^\infty \!\!\!\!
  d\tau \!\!\int_0^\infty \!\!\!\! d\tau' \gamma(t-\tau-\tau'
  )P_1(t-\tau-\tau')_ {\text{asy}}
  w_2(\tau')z_3(\tau)\nonumber.
\end{eqnarray}
$z_i(\tau):=1-\int_0^\tau d\tau' w_i(\tau')$ denotes  the probability to
wait longer than $\tau$ in state i.
Linearizing the rate $\gamma(t)$ with respect to $\alpha$,
$\gamma(t)=r_0(D)(1-\alpha \cos(\Omega t))$
and making the Ansatz 
$P_1(t)_{\text{asy}}=\sum_{k=-\infty}^\infty p_k \exp(i k \Omega t)$
eventually leads to a tridiagonal recurrence relation for the Fourier coefficients $p_k$.
%\begin{eqnarray}\label{tridiag}  
%  \big(\gamma_{0}(\hat w_2(k\Omega)\hat
%  w_3(k\Omega)-1)-ik\Omega\big)p_k+
%  \qquad\qquad&&\\
%  \gamma_1 \big(\hat w_2(k\Omega)\hat w_3(k\Omega)-1\big)
%  (p_{k+1}+p_{k-1})&=&0\nonumber \\
%   (1\!+\!\gamma_{0}(\bar w_{2}+\bar w_{3}))p_0\!+\!
%\gamma_1(\bar w_{2}+\bar w_{3}) (p_{1}+p_{-1})&=&1,
%\end{eqnarray}
%with $\gamma_0=r_0(D)$,
%$\gamma_{1}=-\gamma_0 \alpha/2$,
Solving this tridiagonal recurrence relation following \cite{risken}
we eventually arrive at
\begin{eqnarray}\label{p0p1}
  p_0&=&\frac{\bar w_1}{\bar w }+O(\alpha^2)\\
%  p_1=p^*_{-1}&=&\!\!\frac{\alpha}{2}\frac{\bar w_1}{\bar w }
%  \frac{\hat w_1(\Omega)[1-\hat w_2(\Omega)\hat w_3(\Omega)]}
%  {1-\hat w_1(\Omega)\hat w_2(\Omega)\hat w_3(\Omega)
%    }+O(\alpha^3)\nonumber,
  p_1=p^*_{-1}&=&\!\!-i\frac{\alpha}{2}
  \frac{(1-\hat w_1(\Omega))(1-\hat w_2(\Omega)\hat w_3(\Omega))}
  {\Omega\bar w(1-\hat w_1(\Omega)\hat w_2(\Omega)\hat w_3(\Omega)
    )}+O(\alpha^3)\nonumber,
\end{eqnarray}
with
\begin{eqnarray}\label{defw}
\hat w_i(\Omega)=\int_0^\infty \!\!\!\!\!d\tau  e^{-i\Omega \tau} w_i(\tau)
\;\mbox{and}\;\;
\bar w_i=\int_0^\infty \!\!\!\!\!d\tau \tau w_i(\tau).
\end{eqnarray}
%\cite{oa03}
where $\bar w=\bar w_1+\bar w_2+\bar w_3$ is the mean time for one cycle
of the undriven system and $\bar w_1$ and $\hat w_1(\Omega)$
are calculated according to eq. (\ref{defw})
with the unperturbed WTD for state 1,
  $w_1(\tau)=\gamma_0\exp(-\gamma_0 \tau)$.
Higher  Fourier coefficients $p_k,\;k\ge 2$ are at least of order $O(\alpha^2)$. 
%For the asymptotic probability to be in state 2 one gets
%\begin{eqnarray*}
%  P_2(t)_{\text{asy}}=\gamma_0 p_0 \bar w_2+2 \Re \big[(\gamma_0 p_1+\gamma_1
%p_0)\hat z_2(\Omega)e^{i \Omega t}\big]+O(\alpha^2)
%\end{eqnarray*}
%with $\hat z_{2}(\Omega)=\int_0^\infty d\tau e^{-i\Omega \tau}
%z_2(\tau)$.
Therefore one arrives at
\begin{eqnarray*}\label{asympoutput}
  \left<x(t)\right>_{\text{asy}}
%&=&x_1 P_2(t)_{\text{asy}}+
%  x_0(1-P_2(t)_{\text{asy}})\\
  &=&\left<x\right>_0+\alpha\frac{(x_1-x_0)}{\Omega \bar w} 
  A \cos (\Omega t+\hat \phi))+O(\alpha^2)
%\nonumber\\
\end{eqnarray*}
where $\left<x\right>_0=\frac{1}{\bar w}
(x_0 (\bar w_1+ \bar w_3)+x_1\bar w_2)$ is the
stationary solution of the process without external signal and
$A=\big|\Gamma(\Omega)\big|$, $\hat \phi=\arg(\Gamma(\Omega))$ and 
\begin{eqnarray}\label{gamma}
\Gamma(\Omega)&=&i\frac{(1-\hat w_1(\Omega))(1-\hat w_2(\Omega))}{1-
  \hat w_1(\Omega) \hat w_2(\Omega)\hat
  w_3(\Omega)}.
\end{eqnarray}
The SPA $\eta$ is the ratio between
the power of the
output $x(t)$ at the frequency of the input signal and the power of the
input signal. It is now given by
\begin{eqnarray}\label{spa}
  \eta&=&
%\frac{A^2}{A_0^2}=
\frac{(x_1-x_0)^2}{ D^2 \Omega^2 \bar w^2}
\big|\Gamma(\Omega)\big|^2+O(\alpha^2).
\end{eqnarray}
Note that not only $\bar w$ depend on the noise
strength $D$ but also $\Gamma(\Omega)$ due to the noise dependence of
$\hat w_1(\Omega)$.

\begin{figure}[htbp]
  \begin{center}
    \includegraphics[scale=0.45]{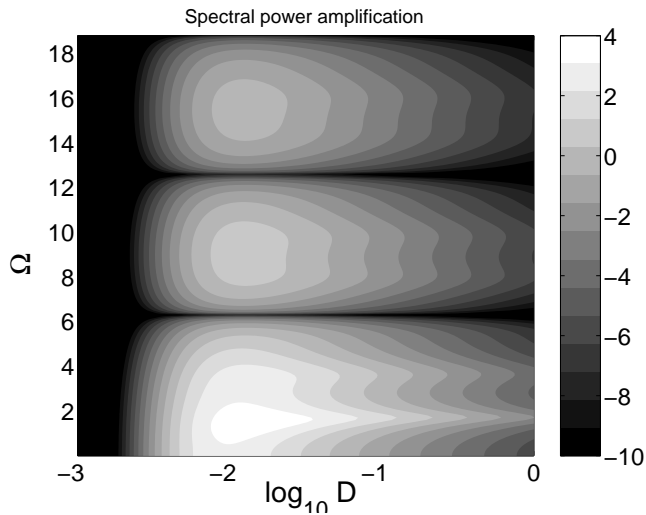}
  \end{center} 
\caption{
    \label{plotspa}
    Spectral power Amplification for fixed firing and refractory times.
    horizontal axes: $\log_{10} D$
    vertical axes: $\Omega $
    Other Parameters: $T_f=1,\; T_r=2,\; \Delta U=0.02,\; x_1=1,\;x_0=0;$
    The minima, where the SNR vanishes, are located at
  $\Omega=2\pi n/T_f, \; n\in N$ The maxima are approximately located
  at $\Omega=2\pi n/(T_f+T_r)$.}
\end{figure}

The SNR is defined as
the ratio between the spectral power at the driving 
frequency and the spectral power density in the
neighborhood of the driving frequency.
In lowest order in $\alpha$ 
this is equal to  the ratio between the spectral power of the driven process
and the spectral power density of the undriven process, both taken at
the driving frequency.
The spectral power density of the process without signal can be 
calculated using a formula from renewal theory \cite{stratonovich}
\begin{eqnarray}\label{spectrum}
  S_0(\Omega)&=&\frac{4(x_1-x_0)^2}{\bar w}
  \frac{\Re G(\Omega)}{\Omega^2}
\end{eqnarray}
with 
\begin{eqnarray}\label{G}
  G(\Omega)=\frac{(1-\hat w_2(\Omega))(1-\hat w_1(\Omega)\hat w_3(\Omega))}
  {1-\hat w_1(\Omega)\hat w_2(\Omega)\hat w_3(\Omega)}.
\end{eqnarray}
Eventually the SNR reads
\begin{eqnarray}\label{snr}
  \text{SNR}
%&=&\frac{\pi A^2}{\lim_{\omega\to \Omega} S(\omega)}\\
  &=&\pi A_0^2 \frac{\eta}{S_0(\Omega)}+O(\alpha^4)\\
  &=&\frac{\pi \alpha^2}{4 \bar w}
  \Big|\frac{1-\hat w_1(\Omega)}{1-\hat w_1(\Omega)\hat w_3(\Omega)}\Big|^2
  \frac{|G(\Omega)|^2}{\Re G(\Omega)}
+O(\alpha^4).\nonumber
\end{eqnarray}
For a fixed firing and refractory time, 
which is a good approximation for an
excitable stochastic FitzHugh-Nagumo system in the low noise regime 
we have $w_2(\tau)=\delta(\tau-T_f)$ and $w_3(\tau)=\delta(\tau-T_r)$ and therefore
$\hat w_2(\Omega)=\exp(-i \Omega T_f)$ and
$\hat w_3(\Omega)=\exp(-i \Omega T_r)$.
This leads to a signal to noise ratio 
%SNR=$\pi\alpha^2/(2 \bar w)$ 
which is independent of the
driving frequency,
\begin{eqnarray}\label{snrdet}
\text{SNR}=\frac{\pi\alpha^2}{2 \bar w}
\end{eqnarray}
%Therein  $\bar w=\frac{1}{\gamma_0}+T_f+T_r$ is the mean time for one
%cycle of the unperturbed system.
In this case, the SPA is proportional to the
spectrum of the unperturbed process,
\begin{eqnarray}\label{etadet}
\lefteqn{  \eta=\frac{\text{SNR}}{\pi A_0^2}S_0(\Omega)
  =\frac{1}{ D^2}\frac{(x_1-x_0)^2}{\bar w^2}}\\
  &&\frac{2 \sin^2
  \frac{\Omega
    T_f}{2}}{\frac{\Omega^2}{2}+\gamma_0^2\big(1-\cos(\Omega
  (T_f+T_r))\big)
  +\gamma_0 \Omega \sin(\Omega (T_f+T_r))}\nonumber.
\end{eqnarray}
with proportionality constant $ (2 D^2 \bar w^2)^{-1}$.
Taking into account the dependence of $\gamma_0$ and therefore
of $\bar w$ on $D$, this function
has a maximum at a finite value of $D$  (see Fig. \ref{plotspa})
which is characteristic for SR.
However also as a function of the driving frequency
the SPA shows maxima at certain frequencies, which has also been
observed in bistable systems with delay \cite{pikovsky}.
%The SPA $\eta$ is 0 if the driving frequency 
%$\Omega/(2 \pi)$ is a multiple of $1/T_f$.
In the Markovian case, where all transitions are rate processes, i.e.
$w_2(\tau)=\gamma_2 \exp(-\gamma_2 \tau)$ and 
$w_3(\tau)=\gamma_3 \exp(-\gamma_3 \tau)$,
we get
% and therefore
%$\hat w_2(\Omega)=\gamma_2/(\gamma_2+i \Omega)$ and
%$\hat w_3(\Omega)=\gamma_3/(\gamma_3+i \Omega)$ the SNR reads
\begin{eqnarray}\label{snrpoisson}
  \text{SNR}=\frac{\pi \alpha^2}{4 \bar w} \frac{\gamma_3^2 +\Omega^2}
  {\gamma_0^2+\gamma_0 \gamma_3+ \gamma_3^2 +\Omega^2}
\end{eqnarray}
This SNR depends on the driving frequency $\Omega$ 
in a monotone way.
However the SPA
\begin{eqnarray}
  \eta=
 \frac{(x_1-x_0)^2(\gamma_3^2+\Omega^2)}{D^2 \bar w^2[(\gamma_0\gamma_2+
    \gamma_0\gamma_3+\gamma_3\gamma_2)^2+
    (\gamma_0^2+\gamma_2^2+\gamma_3^2)\Omega^2+ \Omega^4]}\nonumber
\end{eqnarray}
can show a maximum with respect to the driving frequency.
A reduction to two states, neglecting state 3 by choosing
$w_3(\tau)=\delta(\tau)$ 
gives a frequency independent $\text{SNR}=\frac{\pi \alpha^2}{4 \bar
  w}$ and a SPA 
$\eta=(x_1-x_0)^2/(D^2\bar w^2 [(\gamma_0+\gamma_2)^2+\Omega^2]$
which depends monotonously on the frequency.
For $\gamma_2=\gamma_0$ this corresponds to the situation of
SR in a bistable potential, except that the rate for
the transition back from state 2 to state 1 is not modulated by the
external signal, which leads to an additional factor $1/4$ in the SNR
compared to the well known expression $\text{SNR}=\frac{\pi
  \alpha^2}{2}\gamma_0$ for the signal to noise ratio in a two state
model of a bistable system\cite{gammaitoni}.

%\section{Inter-spike Interval Distribution}
Another quantity, which shows the influence of noise and external
driving in an excitable system
is the inter-spike interval distribution (ISID) W(T). 
In the case of bistable stochastic systems an analogous quantity,
namely the distribution of times for a transition from one stable
state to the other and back again, have been calculated in 
\cite{choi}, \cite{zhou}, \cite{loefstedt}.
 An inter-spike interval is defined as the time between two subsequent
firing events, which  in our model  corresponds to 
two subsequent events of entering state 2.
We calculate this distribution in the limiting case   $A_0/D\to 0$.
The ISID can be written as
\begin{eqnarray}\label{isid}
  W(T)=\int_0^{2 \pi} d\phi W(T|\phi) p(\phi).
\end{eqnarray}
where  $W(T|\phi)$ is the ISID conditioned by the
phase of the external signal at the time of entering state 2
and  $p(\phi)$ is the distribution of signal 
phases at the moment  the system enters state 2.

The conditioned ISID  $W(T|\phi)$ is given by
\begin{eqnarray}\label{defcondisid}
  W(T|\phi)&=&\int_0^T dt (w_2 \circ w_3)(t) w_1(T-t|\phi+t \Omega),
\end{eqnarray}
where $w_1(\tau|\phi)$ is the WTD in state 1
given that the phase at the moment of entering this state was $\phi$.
This distribution reads
$w_1(\tau|\phi)=\gamma_{\phi}(\tau) \exp(-\int_0^\tau dt \gamma_{\phi}(t))$
with
$\gamma_{\phi}(\tau)=r_0(D) \exp(-\alpha \cos(\Omega \tau+\phi))$.

The phase distribution $p(\phi)$ can be readily calculated from the asymptotic
solution $P_1(t)_{\text{asy}}$. It is proportional 
to the probability current from 1 to 2 at the corresponding
time $t=\phi/\Omega$, which is
given by $f(t)=\gamma(t) P_1(t)_{\text{asy}}$. 
The proportionality constant is 
fixed by normalization,
\begin{eqnarray}\label{defpphi}
  p(\phi)=\frac{f(\frac{\phi}{\Omega})}{Z}
  ,\quad Z= \int_0^{2 \pi} d\phi f(\frac{\phi}{\Omega})=\frac{2
  \pi }{\bar w} +O(\alpha^2).
\end{eqnarray}
The probability current $f(t)$ integrated over one period of the external
signal gives the average number of entrances into state 2 during one
period of the external signal \cite{talkner}. 
As a necessary condition for frequency synchronization with the
external signal 
this number has to be one, which leads to the time scale matching
condition $\bar w= \frac{2 \pi }{\Omega}$.

Completing the calculation of $p(\phi)$, we  insert $P_1(t)_{\text{asy}}$ 
using eqs. (\ref{p0p1}) into eq. (\ref{defpphi}) and arrive at 
\begin{eqnarray}\label{phasedistribution}
  p(\phi)&=& \frac{1}{2 \pi}\big( 1-\alpha A \cos(\phi +\hat \phi)\big)+
    O(\alpha^2).
\end{eqnarray}
with  $A=\left|g(\Omega)\right|$, $\hat \phi=\arg(g(\Omega))$
and
\begin{eqnarray}\label{g}
  g(\Omega)&=&\frac{1-w_1(\Omega)}{1-w_1(\Omega)w_2(\Omega)w_3(\Omega)}.
\end{eqnarray}
The first order in $\alpha$ of $p(\phi)$ is sufficient 
to calculate the WTD $W(T)$  
according to eq.~(\ref{isid}) up to order $O(\alpha^2)$.
This can be seen by writing down the general form of $p(\phi)$ up to 
order $O(\alpha)^2$, $p(\phi)=\frac{1}{2\pi}+\alpha f_1(\phi)+
  \alpha^2 f_2(\phi)$.
Due to the normalization the integral over $\phi$ from 0 to $2\pi$ of 
$f_1$ and $f_2$ must vanish.
Inspecting Eq. (\ref{isid}) we notice that the term of order 
$O(\alpha^2)$ which stems
from $O(\alpha^2)$ of $p(\phi)$ and $O(\alpha^0)$ of $W(T|\phi)$
vanishes as  $W(T|\phi)$ does not depend on $\phi$ in 
$O(\alpha^0)$.
Inserting eq. (\ref{defcondisid}) and 
eqs. (\ref{phasedistribution}) into  (\ref{isid}) one can at least in
principle calculate the inter-spike interval distribution for
arbitrary WTD  $w_2$ and $w_3$.

\begin{figure}[htb]
  \begin{center}
    \includegraphics[scale=0.45]{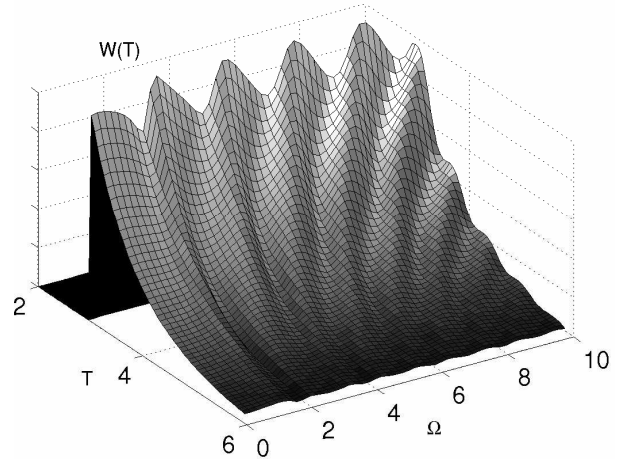}
\end{center}
\caption{ \label{restimesfig}
  Inter spike interval distribution $W(T)$ as a function of the driving
  frequency $\Omega$. $T_f=1,\; T_r=2,\; r_0(D)=1,\; \alpha=0.5$
}
\end{figure}

As an example we explicitly calculate the ISID for  fixed waiting
times $T_f$ and $T_r$ in state 2 and 3 respectively which is
\begin{eqnarray}
\qquad W(T)&=& r_0 e^{-r_0
    (T-T_f-T_r)(1+\frac{\alpha^2}{4})}\Big[1+\frac{\alpha^2}{4}
  \big(\frac{1}{1-2f}+\nonumber\\&&\hspace{-1cm}
2(1+\beta^2)(1+\cos(\Omega T+\phi)\big)\Big] +O(\alpha^4)
\end{eqnarray}
if $T>T_r+T_f$ and 0 otherwise,
$\tan \phi=2fg/(f^2-g^2)$,
$f=1+\beta^2(1-\cos \Delta \phi)+\beta \sin \Delta \phi$ and
$g=\beta \cos \Delta \phi +\beta^2 \sin \Delta \phi$, $\Delta
\phi=\Omega(T_r+T_f)$. 
A plot of the resulting ISID is shown in Fig. \ref{restimesfig}.
The probability that the interval between two subsequent spikes is 
less than $T_f+T_r$ is zero. The distribution for larger
intervals is an exponential decay modulated by the external signal,
which leads to small peaks at  $T=2 \pi n/\Omega$. 
As the calculation is limited to small $\alpha$, which corresponds to
a weak modulation of the transition rate from 1 to 2 due to the
external signal, those peaks are poorly pronounced. 

In conclusion we have derived analytical 
expression for the spectral power amplification and
the signal to noise ratio for small signal amplitudes in a 
non Markovian three state model for stochastic excitable systems. 
The SNR and SPA show a
maximum for a certain finite value of the noise, which indicates
SR. 
However, for many choices of the WTDs in state 2 and 3, 
the SPA as well as the SNR  
show one or several maxima  also as a function of the driving
frequency, i.e. a  ``bona
fide'' resonance. 
In the special case of fixed waiting times in state 2 and 3 
the SNR does not depend on the frequency, indicating, that all  signals are equally well separated from the noisy background,
independent of their frequency.
We have further derived expressions for the inter-spike interval
distribution for small signal amplitudes.

This work was supported by DFG-Sfb 555.
We thank J. Freund for help and fruitful comments.

\end{document}